\documentclass[aps,prl,twocolumn]{revtex4-1}

\usepackage{graphicx}  
\graphicspath{ {images/} } 
\usepackage{latexsym}
\usepackage{slashed}
\usepackage{dcolumn}   
\usepackage{bm}        
\usepackage{amsmath}
\usepackage{amssymb}   
\usepackage{longtable}
\usepackage{float}
\newcommand{\be}{\begin{equation}}
\newcommand{\ee}{\end{equation}}
\newcommand{\bea}{\begin{eqnarray}}
\newcommand{\eea}{\end{eqnarray}}
\newcommand{\bma}{\begin{matrix}}
\newcommand{\ema}{\end{matrix}}

\newcommand{\bml}{\begin{mathletters}}
\newcommand{\eml}{\end{mathletters}}
\newcommand{\bes}{\begin{subequations}}
\newcommand{\ees}{\end{subequations}}

\newcommand{\bi}{\begin{itemize}}
\newcommand{\ei}{\end{itemize}}

\newcommand{\mev}{~{\rm MeV}}

\begin{document}
\title{All in the Family: the quintessential kinship between Inflation and Dark Energy}
\author{P. Q. Hung}
\email{pqh@virginia.edu}
\affiliation{Department of Physics, University of Virginia,
Charlottesville, VA 22904-4714, USA}

\date{\today}

\begin{abstract}
A unified dynamical model of dark energy and inflation is presented, in which both phenomena are driven by axion-like fields-quintessences-of spontaneously broken global $U(1)_A$'s symmetries whose potentials are induced by instantons of the QCD gauge group $SU(3)_c$ for inflation and of a new strongly interacting gauge group $SU(2)_Z$ for dark energy. It is shown that $SU(3)_c$ and $SU(2)_Z$ fit snugly into a unified gauge group $SU(5)_Z$, {\em Ischyr$\acute{o}$s Unification Theory} or {\em IUT}, which is spontaneously broken down to $SU(3)_c \times SU(2)_Z \times U(1)_X$. A judicious choice of $SU(5)_Z$ representations leads to the $SU(3)_c$ and $SU(2)_Z$ couplings growing strong at $\Lambda_{QCD} \sim 200 \mev$ and $\Lambda_Z \sim 10^{-3} eV$ respectively. The model predicts particles carrying $SU(2)_Z$ quantum numbers which can be searched for at colliders such as the LHC and which, as a result, might indirectly reveal the nature of dark energy and perhaps inflation in a laboratory. In addition, the fermionic spectrum of the model contains a possible candidate for dark matter.

\end{abstract}

\maketitle

It is by now almost repetitive to ascertain the importance in our attempts to understand two of the most fundamental mysteries in cosmology: the horizon and flatness problems which gave rise to the elegant idea of Inflation \cite{guth} and the observed phenomenon of an accelerating universe \cite{DE} whose most ``common explanation'' is the assumed existence of some form of ``Dark Energy" which makes up for $\sim 69.4\%$ of the present total energy density of the Universe. Although our understanding of these conundrums is very far from being satisfactory, there seems to be an overwhelming agreement that these two phenomena, albeit occurring at very different epochs and very different energy scales, are driven by some form of ``vacuum energies". This dominance of vacuum energies is what gave rise to the exponential expansion of the very early universe in the form of Inflation and the accelerated expansion of the present universe in the form of Dark Energy. One cannot fail but recognize the similarity between these two fundamental phenomena. It is then reasonable to ask whether or not they came from the {\em same "mother" source} whose two "offsprings" are responsible for Inflation and our current accelerating universe. In other words, could they be the big and little ``siblings" of the same family? Could there be something like an {\em Ischyr$\acute{o}$s Unification Theory} as mentioned in the abstract?. (Ischyr$\acute{o}$s is a Greek adjective for strong.)

From hereon, we will assume that the vacuum energy associated with the accelerating universe comes from a new interaction which grows strong at a scale $\sim 10^{-3} eV$ and NOT from some unknown cancellation among the energies of various vacua. 

If one is willing to accept the aforementioned premise, a clear rationale is in order for the presumed existence of such a {\em Family} and the dynamical reasons that distinguish Inflation from Dark Energy. Where did the vacuum energies come from? Are they constant or do they vary with time-as a manifestation of the time evolution of some quintessence field- during the concerned epochs? What drove Inflation? What drives the present accelerating universe? In what follows, we will focus on two particular models which bring out the kind of kinship referred to in the abstract. (Generalizations could however be contemplated but are beyond the scope of this paper.) In particular, during the time-evolution of the scalar field (inflaton, quintessence), the vacuum energy reflects the energy difference between the so-called false and true vacua. 

The aforementioned ``siblings" refer to particular models of Inflation and Dark Energy whose driving agents are axion-like fields and whose instanton-induced potentials are obtained from confining gauge theories such as $SU(3)_c$ for Inflation \cite{freese} (Chain Inflation) and $SU(2)_Z$ for Dark Energy \cite{hung}. The inflaton field is a pseudo-Nambu-Goldstone (PNG) of a global Peccei-Quinn (PQ) $U(1)_A$ symmetry \cite{PQ}. In a similar manner, the quintessence field for Dark Energy is also a PNG of a global $U(1)_A$ symmetry of the DE model.

In this paper, we will focus primarily on the Dark Energy aspect of the aforementioned {\em IUT}. Details of Chain Inflation based on $SU(3)_c$ are given in Ref.~\cite{freese}.We will concentrate on the conditions for the unification of QCD with the aforementioned $SU(2)_Z$ followed by possible experimental signatures at the LHC and beyond. 

The simplest IUT gauge group is $SU(5)_Z$ (NOT THE SAME as the GUT $SU(5)$) which is vector-like as with $SU(3)_c$ and $SU(2)_Z$. We will assume the following pattern of symmetry breaking:

a) $SU(5)_Z \rightarrow SU(3)_c \times SU(2)_Z \times U(1)_X$ via a 24-dimensional Higgs field of $SU(5)_Z$, $\Phi_{24}$, at a scale $M_S$,

b) $SU(3)_c \times SU(2)_Z \times U(1)_X \rightarrow SU(3)_c \times SU(2)_Z$ via a 10-dimensional Higgs field of $SU(5)_Z$, $\Phi_{10}$, at a scale $M_X$.

From the above pattern, $SU(3)_c$ and $SU(2)_Z$ are unbroken below $M_S$. Although it is beyond the scope of the paper to discuss details of the symmetry breaking, a few remarks are in order concerning the aforementioned Higgs fields. Under $SU(3)_c \times SU(2)_Z \times U(1)_X$, they transform as

i) $\Phi_{24} = (1,1)_0 + (1,3)_0 + (3,2)_{-5} + (\bar{3},2)_5 + (8,1)_0$,

ii) $\Phi_{10} = (1,1)_6 + (\bar{3},1)_{-4} + (3,2)_1$,

where the subscripts in the above expressions represent $U(1)_X$ quantum numbers. From the above decomposition, it is straightforward to see that, in order to achieve the desired pattern of symmetry breaking, one must have $\langle \Phi_{24} \rangle = \langle (1,1)_0 \rangle \neq 0$ and $\langle \Phi_{10} \rangle = \langle (1,1)_6 \rangle \neq 0$ respectively. 

We now turn to the evolution of the $SU(3)_c$ and $SU(2)_Z$ couplings down from their unification scale $M_S$. First,
\be
\label{equal}
\alpha_3(M_S)=\alpha_Z(M_S).
\ee 
Second, left- and right-handed fermions can simply be put in the fundamental representation of $SU(5)_Z$ as follows
\be
\label{fundamental}
\bm{5}=(3,1)_{-2} + (1,2)_3,
\ee 
where $(3,1)_{-2}$ represents the Standard Model (SM) quarks, $q$, while $(1,2)_3$ represents the Dark quarks $Q_Z$. This fundamental representation can now be written in terms of quark fields as
\be
\label{quark}
\Psi= \left( \begin{array}{c}
q \\
Q_Z
\end{array} \right)
\ee
(Not shown in (\ref{fundamental}) are the electroweak quantum numbers.) Notice that, within the framework of our model, the SM and Dark quarks, $q$ and $Q_Z$, are connected to each other via the $SU(5)_Z/SU(3)_c \times SU(2)_Z \times U(1)_X$  gauge bosons. Furthermore, since there are three known generations of SM quarks, there will also be three generations of Dark quarks with the same electroweak quantum numbers. As shown below, an additional set of fermions will be needed to obtain the desired evolution of the couplings, in particular for $\alpha_Z(\Lambda_{Z}) \sim O(1)$, with $\Lambda_{Z} \sim 10^{-3}eV$, as well as for anomaly cancellation in the electroweak sector. It will be seen below that this extra set of fermions contains a possible candidate for dark matter.

To simplify our discussion, we will use the one-loop approximation to the renormalization group equation (neglecting possible scalar contributions for the present purpose):
\be
\frac{1}{\alpha_{i}(M_S)}=\frac{1}{\alpha_{i}(M_Z )} + \frac{\beta^{i}_0}{2 \pi} \ln (\frac{M_S}{M_Z}),
\ee
where $i=3$ for $SU(3)_c$ and $i=Z$ for $SU(2)_Z$ with $M_S$ and $M_Z$ being the IUT mass and Z-boson mass respectively. Taking into account Eq.~(\ref{equal}), we now relate $\alpha_{Z}(M_Z)$ to $\alpha_{3}(M_Z)$ as follows
\be
\label{prediction}
\frac{1}{\alpha_{Z}(M_Z)}=\frac{1}{\alpha_{3}(M_Z )} + \frac{(\beta^{(3)}_0 - \beta^{(Z)}_0)}{2 \pi} \ln (\frac{M_S}{M_Z}).
\ee
Eq.~(\ref{prediction}) is the centerpiece of our predictions for $\alpha_{Z}(M_Z)$ given the experimental value for
\be
\label{alphastrong}
\alpha_{3}(M_Z)=0.1179 \pm 0.0010 .
\ee
In running the couplings from $M_S$ to $M_Z$, we will assume that all non-SM fermions (i.e. not SM quarks and leptons) have masses above $M_Z$ and decouple below it. There are several reasons for making this assumption (as we shall see more explicitly below): these fermions carry SM quantum numbers and would provide unwanted contributions to the evolution of the QCD and electromagnetic couplings. With this constraint in mind, the evolution of the $SU(2)_Z$ coupling from $M_Z$ to $\Lambda_{DE}$ involves only the $SU2)_Z$ gauge fields. We obtain
\be
\label{alphaZ}
\frac{1}{\alpha_{Z}(\Lambda_{DE})}=\frac{1}{\alpha_{Z}(M_Z )} - \frac{11}{3 \pi} \ln (\frac{M_Z}{\Lambda_{DE}}).
\ee
We require $\alpha_{Z}(\Lambda_{Z}) \sim O(1)$ at $\Lambda_{Z} \sim 10^{-3} eV$. Assuming $M_Z/ \Lambda_{Z} = 10^{14}$, we obtain from Eq~(\ref{alphaZ})
\be
\label{constraint}
\alpha_Z(M_Z) =0.026 .
\ee
The above value $0.026$ is our constraint to be satisfied.

To see what additional vector-like fermions will be needed, let us start with just the fundamental representation (\ref{fundamental}). For 3 generations, $\beta^{(3)}_0=(33-12)/3$ and $\beta^{(Z)}_0=(22-12)/3$ so that $\beta^{3}_0 - \beta^{Z}_0 =11/3$, independent of the number of generations (a well-known feature of GUT schemes). Using Eq.~(\ref{prediction}), we obtain $\alpha_Z(M_Z) =0.0367$, in contradiction with the constraint (\ref{constraint}). Additional $SU(5)_Z$ representations are needed.

A search through representations above $\bm{5}$ reveals a surprising match: a single, vector-like, electroweak singlet  $\bm{15}$ of $SU(5)_Z$. We have
\be
\label{15}
\bm{15}=(1,3)_6 + (3,2)_1 + (6,1)_{-4}.
\ee
We shall see below that $(1,3)_6$ could be a dark matter candidate which interacts with SM quarks via the $U(1)_X$ gauge boson with mass $\sim M_X$. We first turn our attention to the RG prediction for $\alpha_Z(M_Z)$. With the addition of $\bm{15}$ to $\beta_0$'s, we obtain $\beta^{(3)}_0=(33-12-4-10)/3$ and $\beta^{(Z)}_0=(22-12-8-12)/3$, where $4/3$, $10/3$, $8/3$ and $12/3$ represent the contributions from $(3,2)$, $(6,1)$, $(1,3)$ and $(3,2)$ respectively. From Eq.~(\ref{prediction}), we obtain the prediction
\be
\label{prediction2}
\alpha_Z(M_Z) =0.0266 .
\ee
This is in excellent agreement with the constraint (\ref{constraint}) and guarantees that $\alpha_{Z}(\Lambda_{Z}) \sim O(1)$ at $\Lambda_{Z} \sim 10^{-3} eV$.

We have covered the model construction aspect of {\em Ischyr$\acute{o}$s Unification Theory} (IUT) above. We now turn to the cosmology aspect of the model, in particular dark energy. We will present a condensed description of aspects of the IUT quintessence, leaving a more complete analysis for a longer version. The main ingredient in the works of \cite{freese} and \cite{hung} was the existence of Peccei-Quinn-like global symmetries, $U(1)_A$, that are spontaneously broken resulting in the existence of a Nambu-Goldstone (NG) boson, the axion, which acquires a mass due to the explicit breaking of the global symmetries by instanton effects of the corresponding strongly interacting gauge theories (QCD for Inflation \cite{freese} and $SU(2)_Z$ for Dark Energy \cite{hung}). In IUT, these global symmetries are $U(1)_{A}^{(3)} \times U(1)_{A}^{(Z)}$, where again the superscripts $3$ and $Z$ are associated with $(3,1)$ and $(1,2)$. The strong interaction energy scales are $\Lambda_{QCD} \sim 200 \mev$ and $\Lambda_{Z} \sim 10^{-3} eV$ and they set the scales for the instanton-induced effective potential of the quintessence axion fields which drive inflation and the accelerating universe respectively.  As described in \cite{freese} and \cite{hung}, the scalar fields containing the aforementioned quintessence fields should be singlets under the strong interaction groups which, in the present context, are $SU(3)_c \times SU(2)_Z$. In addition, they should be of the forms which {\em only couple} to either the SM quarks and/or the Dark quarks but not both at the same time otherwise the respective effective potentials would have the wrong scales. This point is clarified below.

The number of flavors having either $SU(3)_c$ or $SU(2)_Z$ quantum numbers is 6: 3 generations of weak doublets. A Yukawa coupling of $\Psi$ with some $SU(3)_c \times SU(2)_Z$-singlet scalar field would involve a $SU(5)_Z$-singlets, $\phi_0$ and $\tilde{\phi}_0$, and  $SU(5)_Z$-adjoints, $\phi_{24}=(\lambda^a /2)\varphi^a$ and $\tilde{\phi}_{24}=(\lambda^a /2)\tilde{\varphi}^a$ with $a=1,..,24$, since $\bar{5} \times 5 = 1 +24$ and $\bm{24} =  (1,1)_0 + (1,3)_0 + (3,2)_{-5} + (\bar{3},2)_5 + (8,1)_0$ which contains a singlet $(1,1)_0$. Let us now introduce the following scalar fields 
\begin{eqnarray*}
\label{O}
{\mathcal O}_1&=&\phi_{24}-(\frac{1}{\sqrt{15}})\phi_0 \\
{\mathcal O}_2 &=&\tilde{\phi}_{24}+(\frac{3}{2\sqrt{15}})\tilde{\phi}_0 
\end{eqnarray*}
From (\ref{O}), one can see that
\begin{eqnarray*}
\label{operators}
\langle {\mathcal O}_1 \rangle&=& (\frac{\lambda_{12}}{2}) \langle \varphi_{12} \rangle -(\frac{1}{\sqrt{15}}) \langle \varphi_0 \rangle =-\sqrt{\frac{5}{12}} diag(0,0,0,1,1) \langle \varphi \rangle \\
\langle {\mathcal O}_2 \rangle&=& (\frac{\lambda_{12}}{2})\langle \tilde{\varphi}_{12} \rangle +(\frac{3}{2\sqrt{15}}) \langle \tilde{\varphi}_0 \rangle= \sqrt{\frac{5}{12}} diag(1,1,1,0,0) \langle \tilde{\varphi} \rangle,
\end{eqnarray*}
where we have assumed $\langle \varphi_{12} \rangle = \langle \varphi_{0} \rangle = \langle \varphi \rangle$ and $\langle \tilde{\varphi}_{12} \rangle = \langle \tilde{\varphi}_{0} \rangle = \langle \tilde{\varphi} \rangle$. Notice that the aforementioned scalars are all $(1,1)$ singlets of $SU(3)_c \times SU(2)_Z$ so that ${\mathcal O}_1$ and ${\mathcal O}_2$ contain the singlet complex scalar fields, $\sigma_Z$ and $\sigma^{(3)}$, of $ U(1)_{A}^{(Z)}$ and $U(1)_{A}^{(3)}$ respectively. In other words, $\sigma_Z$ couples with $Q_Z$ while $\sigma^{(3)}$ couples with $q$. The Yukawa couplings with the SM and Dark components of $\Psi$ can be written as $\sum_{i=1}^{6} Y^{(3)}_i  \bar{q}_{L,i} \sigma^{(3)} q_{R,i} + H.c.$ and $\sum_{i=1}^{6} Y^{(Z)}_i  \bar{Q^{Z}}_{L,i} \sigma_Z Q^{Z}_{R,i} + H.c.$. We focus here on the dark sector, leaving a more complete discussion for a longer version. With
\be
\label{sigmaz}
\sigma_Z = v_Z \exp(\imath a_Z/v_Z) + \tilde{\sigma}_Z,
\ee
where $\langle  \tilde{\sigma}_Z \rangle =0$ and $\langle  a_Z \rangle =0$. Here, the dark axion $a_Z$, a pseudo Nambu-Goldstone boson, is the quintessence field of our dark energy model. The  $U(1)_{A}^{(Z)}$ symmetry with $Q^{Z}_{L,R} \rightarrow \exp(\mp \imath \alpha_Z) \, Q^{Z}_{L,R}$ and $\sigma_Z \rightarrow \exp(-2 \imath \alpha_Z) \,\sigma_Z$ is explicitly broken by $SU(2)_Z$ instantons. There remains an unbroken $Z_6$ symmetry (for six flavors) which implies six degenerate vacua. The instanton-induced potential for $a_Z$ is $V=\Lambda_{Z}^4 (1- \cos \frac{n \, a_Z}{v_Z})$ with $n=0,..,5$ reflecting the existence of six degenerate vacua. Following \cite{freese} and \cite{hung}, we add a soft-breaking term \cite{sikivie} to lift the vacuum degeneracy of the form $\kappa (a_Z/2 \pi v_Z)$, giving
\be
\label{potential}
V=\Lambda_{Z}^4 (1- \cos \frac{n \, a_Z}{v_Z}) + \kappa \frac{a_Z}{2 \pi v_Z} ,
\ee
with $n=0,..,5$. It is not unreasonable to assume that the vacuum at $a_Z/v_Z=10 \pi$ its vacuum energy is $\sim \Lambda_{Z}^4$ giving roughly $\kappa \sim \Lambda_{Z}^4 /5$. The vacua are in ascending energies, from $0$ (the true vacuum) to $ \Lambda_{Z}^4$. As $T < \Lambda_Z$, the universe can be trapped in any of the false vacua. Even if it were trapped in the lowest-energy ($\Lambda_{Z}^4 /5$) false vacuum at $a_Z/v_Z=2 \pi$, the transition time was estimated in \cite{hung} to be $\tau \geq (10^{-106} s) \exp(10^{62})$. The universe would keep on accelerating for a very long time! 

From the electroweak singlet of (\ref{15}), namely $\bm{15}=(1,3)_6 + (3,2)_1 + (6,1)_{-4}$, one can speculate that the $(1,3)_6$ fermion could be a dark matter candidate which interacts with SM quarks via the heavy $U(1)_X$ gauge boson at tree-level. Since $SU(2)_Z$ gauge group confines at a scale $\sim 10^{-3} eV$, "dark baryons" are {\em bosons} with a size of the order $\hbar c/ 10^{-3} eV \sim 1 \, mm$. The quintessence field, $a_Z$, has a mass squared $\sim \Lambda_{Z}^3/v_{Z}$ ($m_{a_Z} \sim 10^{-10} eV$) and may play the role of an ultralight bosonic dark matter. These issues are under investigation.

We now turn briefly to possible indirect signatures of dark energy at colliders. We start with $Q_Z$ ($(1,2)_3$) of $\bm{5}$. This carries the same electroweak quantum numbers as the SM quarks, meaning $(U_Z(+2/3), D_Z(-1/3))_L$ and $U_{Z,R}(+2/3)$, $D_{Z,R}(-1/3)$.The existence of the dark quarks requires anomaly cancellation in the electroweak sector. We propose two methods: 1) the traditional way by adding leptons; 2) the untraditional way by adding mirror fermions \cite{hung2}. 

1) Traditional method: Unlike SM quarks, the dark quarks have only {\em two dark colors}.It is straightforward to see that we need $ (L_{U}^Z (1/6), L_{Z}^D (-5/6))_L$ and $L_{U,R}^Z(1/6)$, $L_{D,R}^Z (-5/6)$. These leptons with unconventional electric charges could be indirect signals of the dark energy model proposed here. Signatures for dark quarks are more subtle and will be discussed elsewhere. They could be produced in pair at the LHC for example through electroweak processes and have to be distinguished from a heavy lepton.

2) Mirror fermions: These fermions were proposed in a model of non-sterile right-handed neutrinos with electroweak-scale masses \cite{hung2}. For every left-handed SM doublet there is a right-handed doublet of mirror fermions and the same goes for weak singlet fermions. For the SM sector, anomaly cancellation occurs between SM and mirror quarks and leptons {\em separately}. Here, we only need cancellation in the dark quark sector alone by introducing its mirror counterpart, without even a need to introduce dark leptons.

An extensive investigation of searches for the dark-energy-related fermions as well as those of the aforementioned dark matter candidate(s) is in progress and will be presented elsewhere.Various cosmological issues such as the $H0$ tension will also be investigated.

 \begin{acknowledgements}
 
 \end{acknowledgements}


\begin{thebibliography}{50}
\bibitem{guth}
A.~H.~Guth,
Phys. Rev. D \textbf{23}, 347-356 (1981)
doi:10.1103/PhysRevD.23.347;
A.~D.~Linde,
Phys. Lett. B \textbf{108}, 389-393 (1982)
doi:10.1016/0370-2693(82)91219-9;
A.~Albrecht and P.~J.~Steinhardt,
Phys. Rev. Lett. \textbf{48}, 1220-1223 (1982)
doi:10.1103/PhysRevLett.48.1220.
\bibitem{DE}
S.~Perlmutter \textit{et al.} [Supernova Cosmology Project],
Astrophys. J. \textbf{517}, 565-586 (1999)
doi:10.1086/307221
[arXiv:astro-ph/9812133 [astro-ph]].;
A.~G.~Riess \textit{et al.} [Supernova Search Team],
Astron. J. \textbf{116}, 1009-1038 (1998)
doi:10.1086/300499
[arXiv:astro-ph/9805201 [astro-ph]].
\bibitem{freese}
K.~Freese, J.~T.~Liu and D.~Spolyar,
Phys. Rev. D \textbf{72}, 123521 (2005)
doi:10.1103/PhysRevD.72.123521
[arXiv:hep-ph/0502177 [hep-ph]].
\bibitem{hung}
P.~Q.~Hung,
Nucl. Phys. B \textbf{747}, 55-87 (2006)
doi:10.1016/j.nuclphysb.2006.04.021[arXiv:hep-ph/0512282 [hep-ph]]; P.~Q.~Hung,
J. Phys. A \textbf{40}, 6871 (2007)
doi:10.1088/1751-8113/40/25/S32
[arXiv:astro-ph/0612245 [astro-ph]].
\bibitem{PQ}
R.~D.~Peccei and H.~R.~Quinn,
Phys. Rev. Lett. \textbf{38}, 1440-1443 (1977)
doi:10.1103/PhysRevLett.38.1440
\bibitem{sikivie}
P.~Sikivie,
Phys. Rev. Lett. \textbf{48}, 1156-1159 (1982)
doi:10.1103/PhysRevLett.48.1156
\bibitem{hung2}
P.~Q.~Hung,
Phys. Lett. B \textbf{649}, 275-279 (2007)
doi:10.1016/j.physletb.2007.03.067[arXiv:hep-ph/0612004 [hep-ph]].
\end{thebibliography}
\end{document}